\documentstyle[aps,epsf]{revtex}

\begin{document}
\draft

\def\bfs{\bf}

\title
{Theory of Hall Effect and Electrical Transport in High-$T_c$ Cuprates: \\ 
Effects of Antiferromagnetic Spin Fluctuations
}

\author
{ 
Kazuki {\sc Kanki}\footnote{E-mail: kanki@ms.cias.osakafu-u.ac.jp}
and Hiroshi {\sc Kontani}$^{1}$
}

\address
{
College of Integrated Arts and Sciences, Osaka Prefecture University, 
Sakai 599-8531\\
$^1$Institute for Solid State Physics, University of Tokyo, 
7-22-1 Roppongi, Minato-ku, Tokyo 106-8666
}

\date{October 12, 1998}
\sloppy
\maketitle

\begin{abstract}
In the normal state of high-$T_c$ cuprates, 
the Hall coefficient shows remarkable temperature dependence, 
and its absolute value is enhanced 
in comparison with that value simply estimated on the basis of band structure. 
It has been recognized that 
this temperature dependence of the Hall coefficient 
is due to highly anisotropic quasiparticle damping rate on the Fermi surface. 
In this paper 
we further take account of the vertex correction to the current vertex 
arising from quasiparticle interactions. 
Then the transport current is transformed 
to a large extent from the quasiparticle velocity, 
and is no longer proportional to the latter. 
As a consequence some pieces of the Fermi surface 
outside of the antiferromagnetic Brillouin zone 
make negative contribution to the Hall conductivity, 
even if the curvature of the Fermi surface is hole-like.
The Hall coefficient is much larger at low temperatures 
than the estimate made without the vertex correction. 
Temperature dependence of the antiferromagnetic spin correlation length 
is also crucial to cause remarkable temperature dependence 
of the Hall coefficient. 
In our treatment the Hall coefficient of the electron-doped cuprates 
can be negative despite hole-like curvature of the Fermi surface.
\end{abstract}


\section{Introduction}

The anomalous normal state transport properties of high-$T_c$ superconductors 
have attracted much attention for a decade. 
For a wide temperature ($T$) range the electrical resistivity shows $T$-linear 
behavior in the optimally doped and under-doped compounds. 
This behavior is understood as a result of the presence of a second energy 
scale well below the Fermi energy.
Such an energy scale is related\cite{moriya90,kohno91} 
to strong antiferromagnetic spin fluctuations 
which develop due to the large antiferromagnetic coupling between nearest 
neighbor Cu spins via the superexchange mechanism. 
Furthermore the electrical resistivity of samples in a over-doping region 
shows a gradual 
change from $T$-linear behavior to $T^2$-behavior\cite{kubo}, 
and the metallic properties in the over-doping limit 
recover conventional behavior of a Fermi liquid. 

In the present paper our main concern is the Hall coefficient $R_H$. 
It is usually related to charge carrier density $n$ as $R_H^{-1}=nec$, 
and its sign to carrier type (electron- or hole-like) 
in conventional metals and semiconductors. 
If the anisotropy in transport relaxation time $\tau_{\rm tr}$ is neglected, 
conductivity and Hall conductivity 
are proportional to $\tau_{\rm tr}$ and 
$\tau_{\rm tr}^2$, respectively. 
Then Hall coefficient $R_H=\sigma_{xy}/(H\sigma_{xx}\sigma_{yy})$ does not 
depend on $\tau_{\rm tr}$, and shows no significant temperature dependence. 
However if transport relaxation time is dependent on temperature 
differently from point to point 
on the Fermi surface, 
the Hall coefficient can depend on temperature. 
In the normal state of hole-doped cuprate 
superconductors\cite{kubo,chien,harris,xiong,takeda,nishikawa,hwang,jin}, 
the sign of the Hall coefficient is positive, 
and the Hall coefficient 
decreases as the temperature increases for a wide temperature range. 
Hole-like character of the Fermi surface has been observed by 
angle-resolved photoemission spectroscopy (ARPES) 
experiments\cite{dessau,marshall,ding}. 
Moreover in samples with lower doping rate 
the absolute value of the Hall coefficient is larger 
and the temperature dependence of the Hall coefficient is stronger. 
The temperature dependence of the Hall coefficient 
has been recently recognized\cite{lercher,stojkovic,yanase} to be due to 
the strongly anisotropic scattering of quasiparticles 
from antiferromagnetic spin fluctuations. 
However it also has been pointed out that the temperature dependence of 
the anisotropy in the quasiparticle damping 
is insufficient to account fully for the Hall coefficient. 

Temperature dependence of the Hall angle $\theta_H$ also 
has attracted attention\cite{chien,harris,xiong}. 
If transport relaxation time $\tau_{\rm tr}$ is isotropic, 
$\cot\theta_H=\sigma_{xx}/\sigma_{xy}$ is proportional to $\tau_{\rm tr}^{-1}$ 
and therefore should show similar temperature dependence 
as that of resistivity. 
However in high-$T_c$ cuprates $\cot\theta_H$ shows nearly $T^2$-behavior 
rather than the $T$-linear behavior of the resistivity. 
This fact also indicate the existence of anisotropy 
in the transport relaxation time. 

In this paper we consider the electrical conductivity 
and the Hall conductivity 
on the basis of the nearly antiferromagnetic Fermi liquid (NAFL) model 
taking account of the vertex correction to the current vertex 
arising from interactions between quasiparticles. 
Study based on the fluctuation exchange (FLEX) approximation\cite{bickers} 
is reported elsewhere\cite{kontani}. 
Incidentally, Miyake and Narikiyo\cite{miyake} 
has developed a theory, which is supposed to explain 
the anomalous behavior of the Hall coefficient, on the basis of 
some kind of spin fluctuation theory but from different viewpoint 
than ours.

The NAFL model\cite{millis,moriya90} has been used 
to calculate the superconducting transition temperature on the basis of 
the Eliashberg formalism\cite{ueda,monthoux93,moriya94}, 
and to describe magnetic and transport properties\cite{stojkovic,hlubina} 
in the normal state. 
In this model the antiferromagnetic spin fluctuations are provided by giving 
the dynamical spin susceptibility $\chi({\bf q}, \omega)$, 
which is required to provide a quantitative fit to NMR 
or other magnetic experiments. 
Since the dynamical spin susceptibility $\chi({\bf q},\omega)$ peaks 
at the wave vector ${\bf Q}=(\pi,\pi)$ on the simple square lattice 
(the lattice constant is set to be unity), 
quasiparticles located close to those momentum points on the Fermi surface 
which are connected to another point on the Fermi surface 
by ${\bf Q}$ have much larger damping rate 
than that of quasiparticles away from those momentum points, 
and therefore are said\cite{stojkovic,hlubina} 
to be ``hot'' in the literature. 
This distinct difference in quasiparticle lifetimes on the Fermi surface 
leads to the anomalous behavior of the Hall coefficient. 

We base our calculation of conductivities 
on the formalism developed by Eliashberg\cite{eliashberg}, 
which was applied to Hall conductivity 
by Fukuyama, Ebisawa and Wada\cite{fukuyama}, and 
by Kohno and Yamada\cite{kohno88}. 
Eliashberg derived a general formula for conductivity 
on the basis of the Fermi liquid theory starting 
from the finite-temperature current-current correlation function, 
namely the Kubo formula. 
His formula is exact as far as the most singular term 
with respect to the quasiparticle damping rate is concerned; 
in applying to strongly correlated metals 
it is on firmer theoretical grounds 
than is a transport theory 
which makes use of the Boltzmann equation\cite{stojkovic,hlubina}. 

We show that by the vertex correction the current vertex is transformed to 
a large extent from the quasiparticle velocity 
in the presence of highly anisotropic scattering coming from 
strong antiferromagnetic spin correlations. 
As a result the Hall coefficient can be much larger at low temperatures 
than the estimate made without the vertex correction. 
Both the smearing of the Fermi and Bose distribution 
and the temperature dependence 
of the spin correlations cause the unusual temperature dependence 
of the Hall coefficient. 

In an electron-doped cuprate Nd$_{2-x}$Ce$_x$CuO$_{4\pm\delta}$, 
the normal state Hall coefficient 
has been reported\cite{takeda,hagen,jiang,fournier} 
to be negative, 
although there seems to remain some controversy 
in experimental situation\cite{wang,jiang}. 
This negative sign is difficult to understand 
in the framework of quasiparticle transport 
without taking into account the vertex correction to the current vertex, 
since with electron-doping the Fermi surface is expected to remain hole-like. 
The hole-like character of the Fermi surface is predicted\cite{massidda} by 
band structure calculations, and is seen\cite{king,anderson} in 
ARPES experiments. 
In the present paper we show that Hall coefficient can be negative 
due to strong backward scattering between quasiparticles, 
even if the curvature of the Fermi surface is everywhere hole-like. 

In \S 2 we describe expressions of conductivities 
which was derived on the basis of the Fermi liquid theory. 
In \S 3 the nearly antiferromagnetic Fermi liquid model is described, 
and the expression of vertex corrections to the current vertex is given. 
In \S 4 numerical results are shown for the transport current, 
the resistivity and the Hall coefficient. 
Conclusion and some discussions are given in \S 5. 
In the present paper we consider only the transport properties 
in the CuO$_2$-plane, although  
out-of-plane transport phenomena also have attracted 
intense interest\cite{takenaka,hussey96,hussey97,watanabe,hussey98}. 

\section{General Expression for Electrical Conductivities}

A formula for electrical conductivity in interacting electron systems was 
given by Eliashberg\cite{eliashberg} on the basis of the Fermi liquid theory. 
It is exact as far as the term singular with respect to quasiparticle 
damping is concerned. 
The result is given by\cite{note}
\begin{equation}\label{conductivity}
\sigma_{xx}=2e^2\int{d^3 p\over(2\pi)^3}\left\{{1\over 2\gamma_{\bfs p}}
\left(-{\partial f\over\partial\varepsilon}\right)_{\varepsilon=E({\bf p})}
\right\}
v_x^*J_x^*.
\end{equation}
Here the factor two is from the spin degeneracy, and 
$v_\mu^* (\mu=x,y,z)$, 
$E({\bf p})$ and $\gamma_{\bfs p}$ represent the velocity, energy 
and damping rate of a quasiparticle of momentum ${\bf p}$, 
and $f(\varepsilon)$ is the Fermi distribution function. 
The renormalized transport current $J_\mu^*=J_\mu^*({\bf p})$ is related to 
a current vertex function $J_\mu({\bf p}, \varepsilon)$ 
analytically continued with respect to a complex energy variable $\varepsilon$ 
as $J_\mu^*({\bf p})=z_{\bfs p}J_\mu({\bf p}, E({\bf p}))$, 
where $z_{\bfs p}$ is the renormalization factor, 
and the function $J_\mu({\bf p}, \varepsilon)$ satisfies the equation
\begin{equation}\label{vc1}
J_\mu({\bf p}, \varepsilon)=Q_\mu({\bf p}, \varepsilon)+
\int_{-\infty}^\infty{d\varepsilon'\over 4\pi i}
\int{d^3p'\over(2\pi)^3}
{\cal T}_{22}({\bf p}\varepsilon|{\bf p}'\varepsilon')
{2\pi i z_{{\bfs p}'}^2\delta(\varepsilon'-E({\bf p}'))\over 
2i\gamma_{{\bfs p}'}}Q_\mu({\bf p}', \varepsilon'),
\end{equation}
where 
a retarded-advanced Green's function pair ($g_2$-section) is approximated as
\begin{eqnarray}
g_2({\bf p},\varepsilon)&=&
G^{\rm R}({\bf p},\varepsilon)G^{\rm A}({\bf p},\varepsilon)\\
&\cong&{2\pi i z_{{\bfs p}}^2\delta(\varepsilon-E({\bf p}))\over
2i\gamma_{{\bfs p}}}, \label{g2}
\end{eqnarray}
and ${\cal T}_{22}$ is related 
to an analytically continued four-point vertex 
function connected to $g_2$-sections on both sides 
in the particle-hole channel 
as eq. (12) of ref. \cite{eliashberg}. 
The quantity $Q_\mu({\bf p}, \varepsilon)$ is related to 
the quasiparticle velocity $v_\mu^*$ as 
$v_\mu^*=z_{\bfs p}Q_\mu({\bf p}, E({\bf p}))\equiv z_{\bfs p}v_\mu$.
It is expedient to introduce ${\cal T}_{22}^{(0)}$ which does not contain 
a section 2 inside of itself. 
Then we obtain the following equation: 
\begin{equation}
{\cal T}_{22}({\bf p}\varepsilon|{\bf p}'\varepsilon')=
{\cal T}_{22}^{(0)}({\bf p}\varepsilon|{\bf p}'\varepsilon')+
\int_{-\infty}^\infty{d\varepsilon''\over 4\pi i}
\int{d^3p''\over(2\pi)^3}
{\cal T}_{22}^{(0)}({\bf p}\varepsilon|{\bf p}''\varepsilon'')
{2\pi i z_{{\bfs p}''}^2\delta(\varepsilon''-E({\bf p}''))\over
2i\gamma_{{\bfs p}''}}
{\cal T}_{22}({\bf p}''\varepsilon''|{\bf p}'\varepsilon'). 
\end{equation}
Consequently  eq. (\ref{vc1}) can be written in the form:
\begin{equation}\label{j20}
J_\mu({\bf p}, \varepsilon)=Q_\mu({\bf p}, \varepsilon)+
\int_{-\infty}^\infty{d\varepsilon'\over 4\pi i}
\int{d^3p'\over(2\pi)^3}
{\cal T}_{22}^{(0)}({\bf p}\varepsilon|{\bf p}'\varepsilon')
{2\pi i z_{{\bfs p}'}^2\delta(\varepsilon'-E({\bf p}'))\over
2i\gamma_{{\bfs p}'}}J_\mu({\bf p}', \varepsilon').
\end{equation}

The vertex function ${\cal T}_{22}$ in the static limit ($\omega=0$) is 
an imaginary quantity\cite{eliashberg,yamada} 
and is related to the discontinuity 
of the four-point vertex function with respect to 
the energy variables $\varepsilon+\varepsilon'$ and 
$\varepsilon-\varepsilon'$ in the complex plane, 
since the analytic continuation depends on whether the energy variables are 
continued from the upper or the lower half-plane. 
This imaginary quantity is related to the imaginary part of selfenergy, 
and only by treating the correction from this term in a manner 
that preserves the conservation laws 
we can obtain the correct result for the resistivity 
arising from electron-electron interactions\cite{yamada,maebashi}, 
i.e. the resistivity vanishes 
without Umklapp processes. 

A general expression for Hall conductivity in interacting electron systems was 
obtained by Kohno and Yamada\cite{kohno88} 
following an earlier work by Fukuyama, Ebisawa and Wada\cite{fukuyama}. 
It is exact as far as the most singular $(1/\gamma_{\bfs p})^2$-term 
is concerned, 
i.e. it gives the main contribution in the Fermi liquid. 
Their results is given by
\begin{eqnarray}\label{hallc}
\sigma_{xy}&=& 2{e^3\over c}H\int{d^3 p\over(2\pi)^3}
\left[J_x{\partial J_y\over\partial p_y}-{\partial J_x\over\partial p_y}J_y
\right]v_x^*{z_{\bfs p}^2\over(2\gamma_{\bfs p})^2}
\left(-{\partial f\over\partial\varepsilon}\right)_{\varepsilon=E({\bf p})} 
\nonumber \\
&=&
2{e^3\over 2c}H\int{d^3 p\over(2\pi)^3}
\left({\bf J}\times{\partial{\bf J}\over\partial p_\|}\right)_z
|{\bf v}_\perp^*|{z_{\bfs p}^2\over(2\gamma_{\bfs p})^2}
\left(-{\partial f\over\partial\varepsilon}\right)_{\varepsilon=E({\bf p})},
\end{eqnarray}
where $J_\mu=J_\mu({\bf p})=J_\mu({\bf p},E({\bf p}))$. 
Note that the charge of carriers is such as 
$e<0(>0)$ in the electron- (hole-) picture. 
In the first expression it is assumed that $x$- and $y$- directions are 
equivalent. 
In the second expression the momentum integration variables in the plane 
perpendicular to the magnetic field (assumed to be in the $z$-direction) 
are changed\cite{ong} from $(p_x,p_y)$ to 
$(p_\|,p_\perp)$, where $p_\|$ is tangential to the curve of intersection of 
a constant energy surface with a plane perpendicular to the magnetic field 
and $p_\perp$ is perpendicular to a constant energy curve in the plane. 
Eq. (\ref{hallc}) is valid as far as the $1/\gamma_{\bfs p}$-term can be 
neglected in comparison with the $(1/\gamma_{\bfs p})^2$-term. 
In the present paper we neglect the $1/\gamma_{\bfs p}$-term,  
although it can be shown that there really are diagrams which contribute 
to the $1/\gamma_{\bfs p}$-term\cite{kohno88,kohno}. 
It is also assumed that interband effects are neglected. 

If we approximate the last factor in the integrand of (\ref{hallc}) as 
$\delta(E({\bf p}))$, 
where the chemical potential for quasiparticles is set to be zero, 
then integration over $p_\perp$ cancels with the factor 
$|{\bf v}_\perp^*|$ and Hall conductivity is reduced to the form
\begin{equation}\label{hallcfs}
\sigma_{xy}=2
{e^3\over 2c}H\int_{\rm FS}{dp_z dp_\|\over(2\pi)^3}
\left({\bf J}\times{\partial{\bf J}\over\partial p_\|}\right)_z
{z_{\bfs p}^2\over(2\gamma_{\bfs p})^2},
\end{equation}
where the integration is over the Fermi surface. 
Therefore Hall conductivity is given by an integration over $p_z$ 
of the area swept out by $(J_x,J_y)$ 
weighted by a factor proportional to the square of unrenormalized lifetime 
of quasiparticles. 
Moreover if we assume free-electron dispersion, i.e. 
$E({\bf p})={\bf p}^2/2m^*$, $v_\mu^*=p_\mu/m^*$, 
and $J_\mu^*=v_\mu^* C_{\rm U}$, then we obtain the well known Drude formulae: 
$\sigma=ne^2\tau_{\rm tr}/m^*$ and $R_H^{-1}=nec$. 
Here $C_{\rm U}$ is a constant independent of ${\bf p}$, 
$n$ is the charge carrier density 
$p_F^3/(3\pi)^2$ ($p_F$ is the Fermi momentum), 
and $\tau_{\rm tr}=C_{\rm U}/(2\gamma_{\bfs p})$ 
is the transport relaxation time. 
As discussed by Yamada and Yosida\cite{yamada,maebashi}, 
$C_{\rm U}$ is divergent 
and resistivity vanishes 
if we consider 
only electron-electron interactions and do not take Umklapp processes 
into account. 

In an anisotropic system on a lattice with a large Fermi surface, 
the vertex correction arising from quasiparticle interaction (\ref{vc1}) 
cannot be reduced to a momentum independent factor $C_{\rm U}$ 
like that introduced in the last paragraph 
under the assumption of isotropy, 
and ${\bf J}^*$ is not in the same direction of ${\bf v}^*$, 
except when the momentum is at symmetric points 
or at the Brillouin zone boundary. 
Therefore the effect of this vertex correction remains 
to be reflected in Hall coefficient, 
while the factor $C_{\rm U}$ and quasiparticle damping $\gamma$ cancel out 
in Hall coefficient in free-electron-like isotropic approximation. 

In the expressions for conductivities, 
$v_\mu^*$, $J_\mu^*$ and $\gamma_{\bfs p}$ are renormalized by $z_{\bfs p}$, 
and the density of states of quasiparticles 
$\delta(E({\bf p}))$ is enhanced by $z_{\bfs p}^{-1}$. 
Therefore all renormalizations cancel with each other. 
This cancellation applies also to $g_2$-sections given by (\ref{g2}), 
which appear in the expressions for the vertex correction 
to the current vertex.

\section{Vertex Corrections 
in the Nearly Antiferromagnetic Fermi Liquid Model}

We consider in this paper the nearly antiferromagnetic Fermi liquid model 
on the simple square lattice 
to describe the interactions between the quasiparticles in a CuO$_2$ plane. 
The model Hamiltonian is given by 
\begin{equation}
{\cal H}={\cal H}_0+{\cal H}_{\rm int},
\end{equation}
where ${\cal H}_0$ describes the electron dispersion relation, 
which is represented by $\varepsilon({\bf p})$. 
The interaction between electrons is 
described\cite{monthoux93,monthoux97} by
\begin{equation}
{\cal H}_{\rm int}={1\over N}\sum_{\bfs q}\bar{g}({\bf q})
{\bf s}({\bf q})\cdot{\bf S}(-{\bf q}),
\end{equation}
where 
\begin{equation}
{\bf s}({\bf q})={1\over 2}\sum_{{\bfs q}\alpha\beta}
a_{{\bfs k}+{\bfs q}\alpha}^\dagger{\bf\sigma}_{\alpha\beta}
a_{{\bfs k}\beta},
\end{equation}
and ${\bf S}({\bf q})$ is the spin-fluctuation operator. 
For simplicity we will ignore the momentum dependence of the coupling 
between electrons and spin fluctuations and set 
$\bar{g}({\bf q})=\bar{g}$. 
The electron interaction is specified by 
\begin{equation}
V_{\rm eff}({\bf q},\omega)=g^2\chi({\bf q},\omega),
\end{equation}
where $\chi({\bf q},\omega)$ is the dynamical spin susceptibility, 
and $g^2=(3/4)\bar{g}^2$. 
We adopt the form of $\chi({\bf q},\omega)$, 
which has been shown\cite{millis} to provide a quantitative 
fit to the NMR experiments:
\begin{equation}\label{chiqo}
\chi({\bf q},\omega)
={\chi_{\bfs Q}\over 1+\xi^2({\bf q}-{\bf Q})^2-
i(\omega/\omega_{\rm SF})},\ \ \  q_x>0, q_y>0
\end{equation}
where 
\begin{equation}
\chi({\bf q},\omega)=i\int_0^\infty dt e^{i(\omega+i0)t}
\langle[S_{\bfs q}^l(t), S_{-{\bfs q}}^l(0)]\rangle, \ \ \ \ 
(l=x,y,z)
\end{equation}
$\chi_{\bfs Q}$ is the static spin susceptibility at wave vector 
${\bf Q}=(\pi, \pi)$, 
and $\xi$ is the antiferromagnetic spin correlation length. 
This form of the dynamical spin susceptibility was derived\cite{moriya90} 
on the basis of the 
self-consistent renormalization (SCR) theory of spin fluctuations. 
It has been widely 
used\cite{imai,thelen,barzykin,zha,itoh,zheng,magishi95,magishi96} 
to analyze NMR and neutron-scattering experiments.

The quasiparticle damping rate $\gamma_{\bfs p}$ is 
related to the imaginary part 
of the selfenergy ${\mit\Sigma}^{\rm R}({\bf p},\varepsilon)$ as
$\gamma_{\bfs p}=-z_{\bfs p}{\rm Im}
{\mit\Sigma}^{\rm R}({\bf p},\varepsilon=E({\bf p}))
\equiv z_{\bfs p}{\mit \Delta}_{\bfs p}$, 
where 
$z_{\bfs p}=(1-\partial{\mit\Sigma}({\bf p},\varepsilon)/\partial\varepsilon)
|_{\varepsilon=0}$ is the renormalization factor. 
We approximate the imaginary part of the selfenergy by 
the lowest order term with respect to exchange of spin fluctuations as 
\begin{equation}\label{selfenergy}
{\rm Im}{\mit\Sigma}^{\rm R}({\bf p},\varepsilon=0)
=g^2\int{d^2p'\over(2\pi)^2}
{{\rm Im}\chi^{\rm R}({\bf p}-{\bf p}',\mu-\varepsilon({\bf p}'))\over
\sinh\displaystyle{\varepsilon({\bf p}')-\mu\over T}},
\end{equation}
where $\mu$ represents the chemical potential for electrons. 
According to the relation\cite{yamada}, 
\begin{equation}\label{T2}
\int_{-\infty}^\infty d\varepsilon
\left[\coth{\varepsilon\over 2T}-\tanh{\varepsilon\over 2T}\right]
F(\varepsilon)=
\int_{-\infty}^\infty d\varepsilon
{2\over\sinh\displaystyle{\varepsilon\over T}}F(\varepsilon)
\approx F'(0)(\pi T)^2,
\end{equation}
where $F(\varepsilon)$ is a smooth function over the scale 
$\varepsilon\sim T$, 
at low temperatures the quasiparticle damping is proportional to $T^2$. 
At higher temperatures 
and for quasiparticles with momentum ${\bf p}$ 
near the antiferromagnetic Brillouin zone boundary, 
only a small portion in the momentum space 
contribute significantly to the integral (\ref{selfenergy}), and then 
the quasiparticle damping is proportional to $T$, since 
$\sinh((\varepsilon({\bf p}')-\mu)/T)$ can be approximated 
as $(\varepsilon({\bf p}')-\mu)/T$ in the integrand. 
At the ``cold spots'' which are the farthest points 
from the antiferromagnetic Brillouin zone boundary, 
the quasiparticle damping is proportional to $T^2$ up to higher temperatures. 
Quasiparticles with smaller damping rate 
contribute larger to the conductivity. 
Yanase and Yamada\cite{yanase} evaluated 
the crossover temperature $T_{\rm cr}$ 
from $T^2$ to $T$-linear behavior of the resistivity to be proportional to 
$({\mit \Delta}k)^2$, where ${\mit\Delta}k$ 
is the distance between the ``cold spots'' 
and the antiferromagnetic Brillouin zone boundary. 

\begin{figure}
\epsfxsize=60mm
\centerline{\epsffile{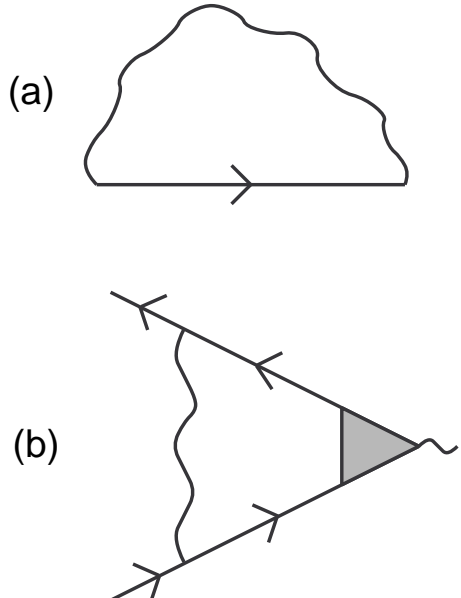}}
\caption{
The selfenergy (a) and the vertex correction (b) in the lowest order 
with respect to exchange of spin fluctuations. 
Straight and wavy lines represent a electron Green's function and 
a fluctuation propagator, respectively.
}
\label{migdal}
\end{figure}

In conformity with the selfenergy represented by fig. \ref{migdal}(a) 
is a vertex correction shown in fig. \ref{migdal}(b). 
Here the electron lines in these figures should be regarded as 
containing the selfenergy correction selfconsistently. 
By the use of (\ref{T2}) we obtain the following formula which is valid only 
at low temperatures: 
\begin{equation}\label{vc22}
J_\mu({\bf p})=v_\mu-g^2\int{d^2p'\over(2\pi)^2}(\pi T)^2
\left.{\partial\over\partial\varepsilon'}{\rm Im}
\chi_{{\bfs p}-{\bfs p}'}^{\rm R}(-\varepsilon')\right|_{\varepsilon'=0}
{z_{{\bfs p}'}^2\delta(E({\bfs p}'))\over 2\gamma_{{\bfs p}'}}
J_\mu({\bf p}'),
\end{equation}
where $J_\mu({\bf p})=J_\mu({\bf p},E({\bf p}))$. 
Since the damping $\gamma_{\bfs p}$ is proportional to $T^2$ 
at low temperatures, 
$J_\mu({\bf p})$ is independent of temperature in this temperature range. 
Eq. (\ref{vc22}) is an integral equation for a function $J_\mu({\bf p})$ 
of momentum ${\bf p}$ on the Fermi surface. 
We solved numerically this equation, and calculated conductivities 
according to eqs. (\ref{conductivity}) and (\ref{hallc}) 
with the use of the resultant $J_\mu({\bf p})$. 

When the antiferromagnetic spin correlations has highly developed, 
i.e. $\chi({\bf q},\omega)$ strongly peaked 
around ${\bf q}={\bf Q}=(\pi,\pi)$, 
the integral in (\ref{vc22}) contribute 
near the ``hot spots'' proportionally to 
$\chi_{\bfs Q}/(\omega_{\rm SF}\xi^2)$, where the factor $\xi^{-2}$ 
comes from the integration with respect to the momentum variable 
in two dimensions. 
If we assume scaling relations for the spin fluctuation parameters 
as $\chi_{\bfs Q}=\alpha\xi^2$ and $\omega_{\rm SF}\xi^2=\Gamma$, then 
$\chi_{\bfs Q}/(\omega_{\rm SF}\xi^2)=\xi^2(\alpha/\Gamma)$. 
Therefore in systems with strong antiferromagnetic spin correlations 
the transport current is transformed to a large extent 
from the quasiparticle velocity, especially near the ``hot spots'', 
and this transformation is sensitive to development of the antiferromagnetic 
spin correlations. 

\begin{figure}
\epsfxsize=60mm
\centerline{\epsffile{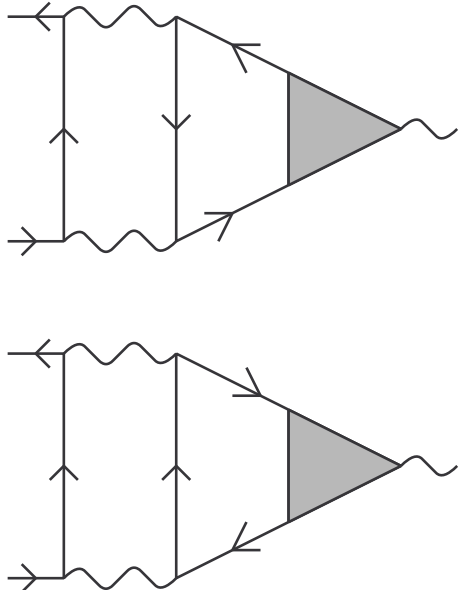}}
\caption{Other diagrams for the vertex correction necessary to strictly 
satisfy the conservation laws in the FLEX approximation.
}
\label{al}
\end{figure}

If the quasiparticles and the spin fluctuations constitute distinct 
degrees of freedom in the system, the vertex correction (\ref{vc22}) 
shown in fig. \ref{migdal}(b) is 
enough to satisfy the local conservation law of electron charge 
in correspondence with the imaginary part of the selfenergy 
represented by fig. \ref{migdal}(a), 
if the selfenergy correction is included selfconsistently. 
However in actuality the spin fluctuations arise from electron-electron 
interactions. 
In the FLEX approximation, 
which is the random phase approximation 
using the Green's function that contains the selfenergy correction 
selfconsistently, 
two more diagrams of the type shown in fig. \ref{al} 
are necessary to strictly satisfy the conservation law.  
These diagrams contribute additionally to (\ref{vc22}) 
at low temperatures as
\begin{equation}\label{vcal}
{4\over 3}g^4\int{d^2p'd^2q\over(2\pi)^4}
|\chi({\bf q},0)|^2\pi\rho_{{\bfs p}-{\bfs q}}(0)
\rho_{{\bfs p}'-{\bfs q}}(0)\rho_{{\bfs p}'}(0)
{(\pi T)^2z_{{\bfs p}'}\over 2\gamma_{{\bfs p}'}}
J_\mu({\bfs p}'), 
\end{equation}
where $\rho_{\bfs p}(0)=z_{\bfs p}\delta(E({\bf p}))$ 
and we have substituted 
for the effective interaction the one coming from 
the dynamical spin susceptibility. 
Similarly to the lowest order diagram (\ref{vc22}), 
the vertex correction (\ref{vcal}) couples strongly a quasiparticles 
at each ``hot spot'' with quasiparticles at other ``hot spots'', 
as well as with itself. 
However to cause strong scattering the difference between ${\bf p}$ and 
${\bf p}'$ does not necessarily be ${\bf Q}$ 
modulo reciprocal lattice vectors, 
since ${\bf p}-{\bf p}'$ does not appear explicitly 
in the expression (\ref{vcal}). 
Therefore these scatterings cancel with each other to some extent. 
The cancellation is complete at ${\bf q}={\bf Q}$. 
Since we assume that $\chi({\bf q},0)$ is strongly peaked 
at ${\bf q}={\bf Q}$, 
we retain only the contribution (\ref{vc22}) 
from the lowest order diagram for ${\cal T}_{22}^{(0)}$ 
in the present paper. 

The vertex correction to the current vertex in the FLEX approximation 
are described in more detail in ref. \cite{kontani},  
where reasons for being able to neglect diagrams of 
the type shown in fig. \ref{al} are fully discussed.

\section{Numerical Results}
\begin{figure}
\epsfxsize=55mm
\centerline{\epsffile{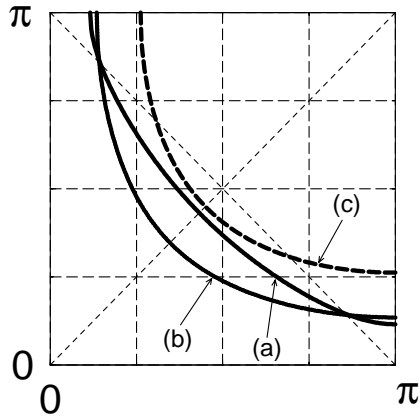}}
\caption{The Fermi surface represented by the dispersion relation with 
(a)$t'=-t/4$ and the chemical potential $\mu=-0.804t$; 
(b)$t'=-0.45t$, $\mu=-1.464t$; 
(c)$t'=-0.45t$, $\mu=-0.584t$ (electron-doped).  
Only a quarter with $p_x>0,p_y>0$ of the first Brillouin zone is shown. }
\label{fs}
\end{figure}

We assume that the electron dispersion relation is given by
\begin{equation}\label{dispersion}
\varepsilon({\bf p})=-2t[\cos(p_x)+\cos(p_y)]-4t'\cos(p_x)\cos(p_y),
\end{equation}
with two hopping matrix elements $t$ and $t'$. 
The lattice constant $a$ of the simple square lattice is set to be unity. 
We choose as examples $t=0.25$ eV, and 
(a) $t'=-t/4$, 
the chemical potential $\mu=-0.804t$, and then 
the electron number density per site is $n=0.90$; 
(b) $t'=-0.45t$, $\mu=-1.464t$ and $n=0.75$. 
The Fermi surfaces are shown in fig. \ref{fs}. 
For simplicity in numerical treatment, we do not determine selfconsistently 
the Fermi surface in the presence of interactions; 
the real part of the selfenergy is fully taken into account 
in ref. \cite{kontani}. 

We adopt as typical parameters for the antiferromagnetic spin correlations 
those used by Monthoux\cite{monthoux97} 
for YBa$_2$Cu$_3$O$_7$.  
The spin fluctuations described by $\chi({\bf q},\omega)$ in the form 
(\ref{chiqo}) are parametrized as 
\begin{eqnarray}\label{sfparam}
\chi_{\bfs Q} &=& \alpha\xi^2, \\ \label{sfparam2}
\omega_{\rm SF}\xi^2 &=& \Gamma
\end{eqnarray}
with scale factors $\alpha=14.7$ eV$^{-1}$ and $\Gamma=0.0732$ eV. 
The temperature dependence of the spin-fluctuation energy $\omega_{\rm SF}$ 
is taken as
$\omega_{\rm SF}=(9.5+4.75[T(K)/100]){\rm meV}$.
We use as the electron-spin-fluctuation coupling constant 
$g^2=0.41$ eV$^2$. 
In our treatment the transport current ${\bf J}({\bf p})$ and 
the Hall coefficient depend neither 
on the momentum independent coupling constant $g$ nor on $\alpha$, 
while the resistivity and the Hall conductivity are proportional 
to $g^2\alpha$ and $(g^2\alpha)^{-2}$ respectively. 
This is because we approximate the selfenergy by the lowest order term 
with respect to exchange of spin fluctuations. 

In the following we measure energy and temperature 
in units of  $4t$, 
and set $|e|=1$ and $c=1$. 

\begin{figure}
\epsfxsize=55mm
\centerline{\epsffile{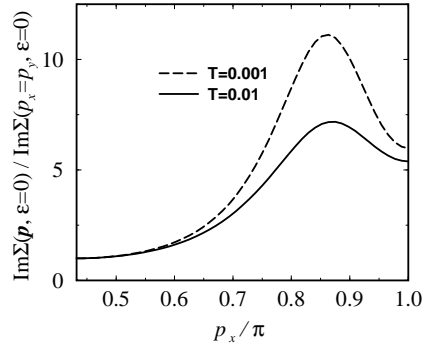}}
\caption{The imaginary part of the selfenergy normalized by the value 
at $p_x=p_y$ as a function of 
one of the components of momentum on the Fermi surface 
at two temperatures $T=0.001$ and $0.01$.
Evaluated for the case (a).
}
\label{gamma}
\end{figure}

In fig. \ref{gamma} is shown for the case (a) the momentum dependence of 
the imaginary part of the selfenergy given by (\ref{selfenergy}) at 
temperatures $T=0.001$ and $0.01$. 
In this figure the values of 
$-{\rm Im}{\mit\Sigma}^{\rm R}({\bf p},\varepsilon=0)$  
normalized by the value at 
$p_x=p_y$ are shown for only a $1/8$-part of the Fermi surface, 
that is, the lower right part of the Fermi surface shown in fig. \ref{fs}. 
The imaginary part of the selfenergy is highly anisotropic and peaks around the 
``hot spots'', each of which is connected to another 
by the wave vector ${\bf Q}$. 
As temperature increases this anisotropy is reduced. 

We approximated $(-\partial f/\partial\varepsilon)_{\varepsilon=E({\bfs p})}$ 
by $\delta(E({\bf p}))$, and calculated the conductivities 
by fixing the quasiparticles on the Fermi surface. 
We performed the following four kinds of numerical evaluation 
for the conductivity and the Hall conductivity. 
\begin{enumerate}
\renewcommand{\labelenumi}{(\roman{enumi})}
\item 
The vertex correction to the current vertex is neglected, 
and the conductivities 
are calculated with the transport current ${\bf J}$ replaced 
by the unrenormalized velocity ${\bf v}$. 
\item The vertex correction (\ref{vc22}) 
is evaluated at the lowest temperature, 
and the resultant transport current ${\bf J}$ is used at all temperatures 
to calculate the conductivities. 
\item 
The values of the spin correlation parameters at each temperature 
to calculate the conductivities 
are used on the evaluation of the vertex correction (\ref{vc22}), 
which is done at the lowest temperature. 
\item
The vertex correction is evaluated at each temperature 
according to the formula (\ref{atan}) given below.  
\end{enumerate}

In order to take account, at least qualitatively, 
of the temperature dependence 
coming from the factor $\coth(\varepsilon'/2T)-\tanh(\varepsilon'/2T)$ 
in the function 
${\cal T}_{22}({\bf p}\varepsilon|{\bf p}'\varepsilon')$ 
in (\ref{j20}), 
in the scheme (iv) 
we solved the following integral equation instead of (\ref{vc22}): 
\begin{equation}\label{atan}
J_\mu({\bf p})=v_\mu+g^2\int{d^2p'\over(2\pi)^2}
{4T\over\Omega({\bf p}-{\bf p}')}
\tan^{-1}\left({\pi^2 T\over 4\Omega({\bf p}-{\bf p}')}\right)
\chi_{\bfs Q}\omega_{\rm SF}
{z_{{\bfs p}'}^2\delta(E({\bf p}'))
\over 2\gamma_{{\bfs p}'}}J_\mu({\bf p}'),
\end{equation}
where $\Omega({\bf q})=\omega_{\rm SF}[1+\xi^2({\bf q}-{\bf Q})^2]$ 
and use is made of the approximate relation: 
\begin{eqnarray}
&&\int_{-\infty}^\infty d\varepsilon
\left[\coth{\varepsilon\over 2T}-\tanh{\varepsilon\over 2T}\right]
{\varepsilon\over\varepsilon^2+\Omega^2}\nonumber\\
&=& {2\pi T\over\Omega}
-2\left[\psi\left({\Omega\over 2\pi T}+1\right)
-\psi\left({\Omega\over 2\pi T}+{1\over 2}\right)\right]\\
&\approx& {4T\over\Omega}\tan^{-1}{\pi^2 T\over 4\Omega},
\end{eqnarray}
where $\psi$ denotes the digamma function. 

\begin{figure}
\epsfxsize=55mm
\centerline{\epsffile{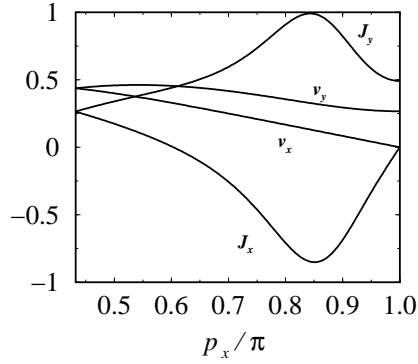}}
\caption{The unrenormalized velocity of the quasiparticles 
and the transport current 
as functions of one of the components of momentum on the Fermi surface. 
Evaluated at $T=0.001$ for the case (a).
}
\label{vj}
\end{figure}

A solution for the case (a) 
of the integral equation (\ref{vc22}) for the vertex correction 
obtained numerically is shown in fig. \ref{vj}. 
We solved the equation at a temperature as low as $10^{-3}$. 
In the temperature region lower than this temperature, 
the solution $J_\mu({\bf p})$ of the equation has only slight 
temperature dependence, 
unless the spin correlation length $\xi$ changes drastically 
in this temperature region. 
At temperatures as high as $10^{-2}$ the equation (\ref{vc22}) no longer 
converges, since the expansion (\ref{T2}) with respect to temperature  
is not valid at high temperatures. 
The numerical results show 
that one of the components of ${\bf J}$ differs in sign 
from the corresponding component of ${\bf v}$, 
and that ${\bf J}$ is enhanced in magnitude from ${\bf v}$, 
near the ``hot spots''. 

\begin{figure}
\epsfxsize=55mm
\centerline{\epsffile{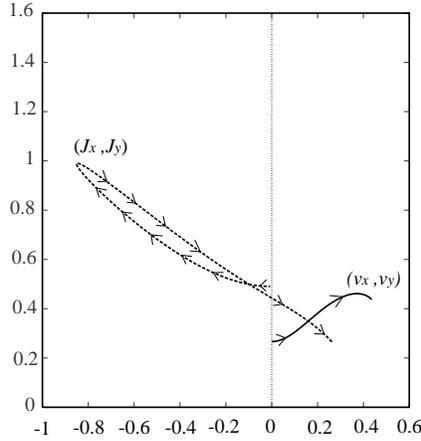}}
\caption{Paths of the unrenormalized velocity ${\bf v}({\bf p})$ 
and the transport current ${\bf J}({\bf p})$ 
in a two-dimensional space 
when the momentum ${\bf p}$ moves along the Fermi surface, 
evaluated for the case (a) at $T=0.001$. 
Of the closed curves only a $1/8$-part corresponding to the lower-right part 
of the Fermi surface is shown. 
The arrows indicate the direction on the paths taken under the integration 
over $p_\|$, that is, 
they denote the direction of the movement on the paths  
when the corresponding momentum moves on the Fermi surface 
from the Brillouin zone boundary where $p_x=\pi$ 
to the diagonal point where $p_x=p_y$.}
\label{vj2d}
\end{figure}

Roughly speaking, in the first Brillouin zone 
the part of the Fermi surface outside of 
the antiferromagnetic Brillouin zone contributes negatively 
to the Hall conductivity. 
In other words, there is in a hole-like Fermi surface a part which 
makes negative contribution to the Hall conductivity. 
On the other hand, 
the whole Fermi surface makes positive contribution 
to the Hall conductivity, if we neglect the vertex correction 
to the current vertex. 
In order to illustrate this fact, in fig. \ref{vj2d} 
the vectors $(J_x, J_y)$ and $(v_x,v_y)$ are plotted in two dimensions. 
According to (\ref{hallcfs}), the values of 
$({\bf J}\times{\rm d}{\bf J})_z$ 
determines the sign of contribution from a segment on the Fermi surface 
to the Hall conductivity. 
Here ${\rm d}{\bf J}$ denotes the variation of ${\bf J}$ along the path 
taken under the integration over $p_\|$. 

The transport current ${\bf J}$ contribute to the conductivities in the form 
${\bf J}({\bf p})z_{\bfs p}/(2\gamma_{\bfs p})=
{\bf J}({\bf p})/(2{\mit \Delta}_{\bfs p})$, 
i.e. in combination with the unrenormalized lifetime of quasiparticles. 
In this combination,  
relative contribution of a part near the first Brillouin zone boundary 
of the Fermi surface to the conductivities is reduced 
because of larger imaginary part of the selfenergy, 
and the overall sign of the Hall conductivity is positive. 

\begin{figure}
\epsfxsize=55mm
\centerline{\epsffile{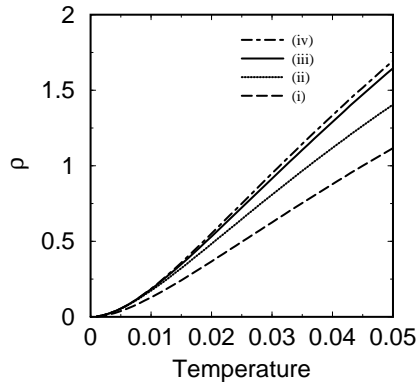}}
\caption{Temperature dependence of the resistivity 
$\rho=\sigma_{xx}^{-1}$ (in units of $\hbar/e^2$) evaluated for the case (a). 
Temperature is in units of $4t$ which is assumed to be 1 eV.
}
\label{resistivity}
\end{figure}

\begin{figure}
\epsfxsize=55mm
\centerline{\epsffile{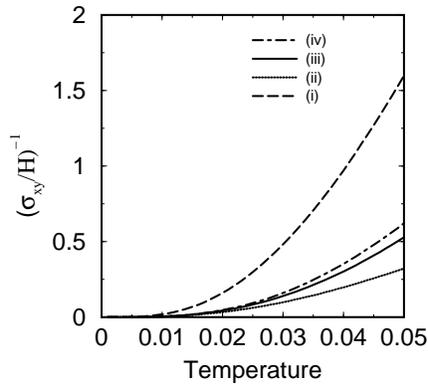}}
\caption{Temperature dependence of the inverse of the Hall conductivity 
$(\sigma_{xy}/H)^{-1}$ (in units of $\hbar^2 c/|e|^3a^2$, 
where $a$ is the lattice constant of the square lattice) 
evaluated for the case (a). 
}
\label{hallcond}
\end{figure}

The temperature dependence of the resistivity $\rho=\sigma_{xx}^{-1}$ and 
the inverse of the Hall conductivity $(\sigma_{xy}/H)^{-1}$ is shown 
for the case (a) 
in figs. \ref{resistivity} and \ref{hallcond} respectively. 
The resistivity is larger when the vertex correction is taken account of. 
This is because the two vectors ${\bf v}$ and ${\bf J}$ point to 
different directions except for those corresponding to momenta 
at symmetric points; 
the two vectors contribute to the conductivity 
in the form of the scalar product ${\bf v}\cdot{\bf J}$. 
Moreover at the point where $p_x=p_y$, 
the absolute value of ${\bf J}$ is reduced from that of ${\bf v}$, 
since the interactions between the quasiparticles lead dominantly 
to backward scattering. 
On the other hand the Hall conductivity is larger when the vertex correction 
is taken account of, 
since the absolute value of ${\bf J}$ is larger than that of ${\bf v}$ 
at most points on the Fermi surface. 

\begin{figure}
\epsfxsize=55mm
\centerline{\epsffile{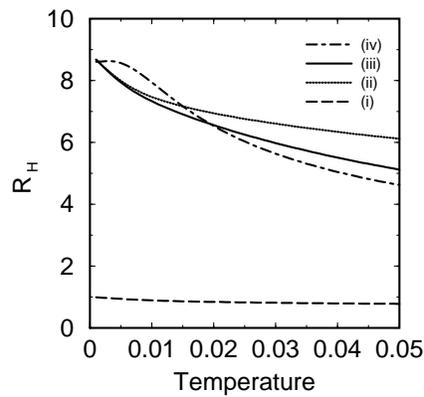}}
\caption{Temperature dependence of the Hall coefficient 
(in units of $a^2/|e|c$) 
evaluated for the case (a). 
}
\label{hallc_temp}
\end{figure}

Temperature dependence of the Hall coefficient is shown 
for the case (a) 
in fig. \ref{hallc_temp}. 
Taking account of the vertex correction leads to nearly an order of magnitude 
larger Hall coefficient at low temperatures. 
The Hall coefficient decreases as the temperature increases. 
This is because quasiparticle damping is strongly anisotropic 
on the Fermi surface, 
and this anisotropy decreases as the temperature increases. 
Moreover the temperature dependence is stronger when the vertex correction is 
taken account of, since then a part of the Fermi surface 
near the first Brillouin zone boundary makes negative contribution 
to the Hall conductivity, 
while at low temperatures part of the Fermi surface near the diagonal points 
(``cold spots'') 
make dominantly positive contribution to the conductivities. 
The temperature dependence of the antiferromagnetic spin correlation length 
also is important in bringing about strong temperature dependence of 
the Hall coefficient; 
as temperature increases and the spin correlation length gets shorter, 
enhancement of the transport current is reduced. 
In the scheme (iv), 
in a narrow low temperature region, effect of the vertex correction 
gets stronger as temperature increases, and then 
decrease of the Hall coefficient with increasing temperature 
is restrained at low temperatures.

\begin{figure}
\epsfxsize=55mm
\centerline{\epsffile{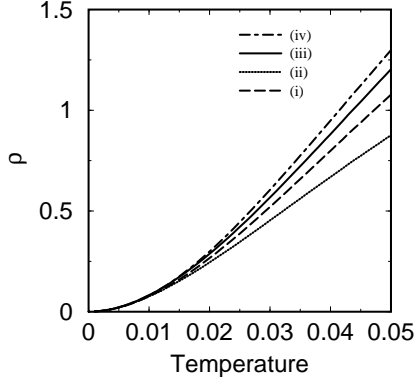}}
\caption{Temperature dependence of the resistivity 
$\rho=\sigma_{xx}^{-1}$ (in units of $\hbar/e^2$) evaluated for the case (b). 
}
\label{resistivity2}
\end{figure}

\begin{figure}
\epsfxsize=55mm
\centerline{\epsffile{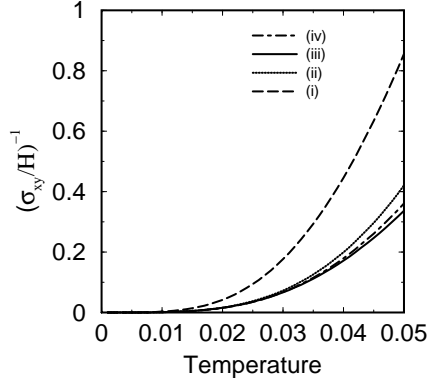}}
\caption{Temperature dependence of the inverse of the Hall conductivity 
$(\sigma_{xy}/H)^{-1}$ (in units of $\hbar^2 c/|e|^3a^2$) 
evaluated for the case (b). 
}
\label{hallcond2}
\end{figure}

\begin{figure}
\epsfxsize=55mm
\centerline{\epsffile{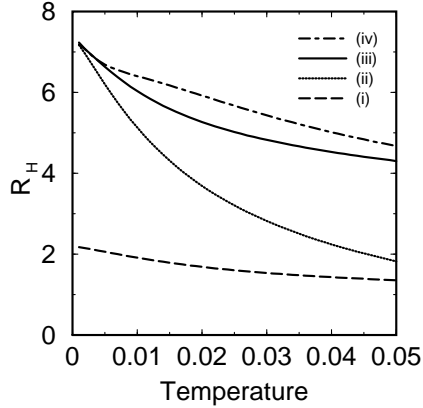}}
\caption{Temperature dependence of the Hall coefficient 
(in units of $a^2/|e|c$) 
evaluated for the case (b). 
}
\label{hallc_temp2}
\end{figure}

In figs. \ref{resistivity2}, \ref{hallcond2} and \ref{hallc_temp2} 
are shown the numerical results for the second example (b); 
see fig. \ref{fs} for the Fermi surface. 
The difference from the case (a) mainly comes from the fact 
that in the case (b) the quasiparticle velocity on the Fermi surface 
near the ``hot spots'' is almost the same in magnitude as that 
near the diagonal point ($p_x=p_y$). 
Since each ``hot spot'' is strongly coupled with another 
in the vertex correction for the current vertex, 
in the case (b) the effect of the vertex correction is stronger 
than that in the case (a). 
Then the transport current is much more transformed and enhanced 
at low temperatures. 
Therefore in the scheme (ii), in which the transport current calculated 
at a low temperature is used to calculate the conductivities at higher 
temperatures, the conductivity $\sigma_{xx}$ is enhance from that estimated 
neglecting the vertex correction. 
This extreme enhancement 
at low temperatures 
also leads to a large temperature dependence 
of the Hall coefficient, 
if the temperature dependence of the transport current is neglected. 
By taking account of the temperature dependence of the antiferromagnetic 
spin correlation length and that coming from the factor 
$\coth(\varepsilon'/2T)-\tanh(\varepsilon'/2T)$ in the integration kernel, 
the degree of transformation and enhancement of the transport current 
are reduced at high temperatures.

\begin{figure}
\epsfxsize=55mm
\centerline{\epsffile{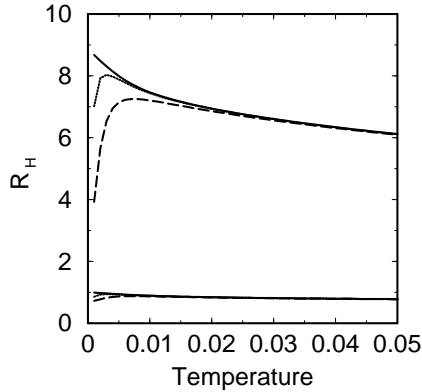}}
\caption{Temperature dependence of the Hall coefficient 
(in units of $a^2/|e|c$) 
in the presence of impurity scattering, which is taken account of 
by giving the imaginary part of the selfenergy a constant 
${\mit \Delta}_{\rm imp}$. 
The upper three curves are calculated in the manner (ii), 
and the lower three curves are calculated in the manner (i). 
For solid, dotted and dashed lines, 
${\mit \Delta}_{\rm imp}=0, 10^{-4},$ and $10^{-3}$ respectively. }
\label{hallc_imp}
\end{figure}

To estimate the effect of impurity scattering 
we calculated the Hall coefficient in the manner (ii)
giving the imaginary part of the selfenergy 
${\mit \Delta}_{\bfs p}=\gamma_{\bfs p}/z_{\bfs p}$ 
in (\ref{conductivity}) and (\ref{hallc}) 
a momentum- and temperature- independent 
constant ${\mit \Delta}_{\rm imp}$ in addition to 
${\mit \Delta}_{\bfs p}^{e-e}$ 
given by (\ref{selfenergy}). 
We show the result in fig. \ref{hallc_imp}. 
At low a temperature region where ${\mit \Delta}_{\bfs p}^{e-e}$ 
becomes smaller than ${\mit \Delta}_{\rm imp}$, 
the Hall coefficient decreases rapidly as the temperature decreases. 
In this calculation 
the vertex correction is given by (\ref{vc22}) with 
${\mit \Delta}_{\bfs p}={\mit \Delta}_{\bfs p}^{e-e}$, 
i.e. we have not treated the vertex correction and the imaginary part of 
selfenergy consistently. 
In spite of this inconsistency in numerical treatment, 
we can expect rapid decrease of the Hall coefficient like the one 
shown in fig. \ref{hallc_imp} 
for the following reason, 
as long as a system remains normal in a temperature regime where 
residual impurity scattering dominates. 
When ${\mit \Delta}_{\rm imp}\gg{\mit \Delta}_{\bfs p}^{e-e}$ 
at low temperatures the vertex correction 
arising from interactions between quasiparticles is negligible, 
since there is a factor proportional to $T^2$ 
arising from the discontinuity of the vertex function in complex energy plane. 
Impurity scattering is expected to be rather isotropic 
compared with scattering from antiferromagnetic spin fluctuations. 
Then the vertex correction from impurity scattering reduces to 
an almost momentum independent factor, and 
the Hall coefficient is approximately given by (\ref{conductivity}) and 
(\ref{hallc}) with $J_\mu$ replaced by $v_\mu$ 
in a temperature region where residual impurity scattering dominates. 

\begin{figure}
\epsfxsize=55mm
\centerline{\epsffile{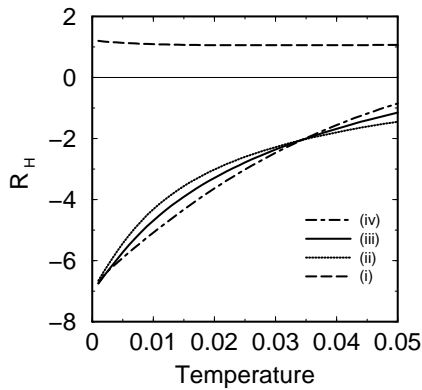}}
\caption{Temperature dependence of the Hall coefficient 
(in units of $a^2/|e|c$) 
in an electron-doped system with $t'=-0.45t,\mu=-0.584t$ and 
the electron number density per cite $n=1.10$. 
}
\label{hallce}
\end{figure}

Now we consider the case of electron-doping. 
There is no essential difference between hole-doped and electron-doped 
systems with regard to the shape of the Fermi surface, that is, 
the Fermi surface is closed around the corner of the Brillouin zone 
$(\pi,\pi)$, except for 
La$_{2-x}$Sr$_x$CuO$_4$ in a over-doping region\cite{ino}. 
Therefore the Hall conductivity never becomes negative in electron-doped 
systems, if the vertex correction to the current vertex is neglected. 
However the Hall conductivity can be negative in our treatment, 
since the greater part of the Fermi surface in the first Brillouin zone 
is outside of the antiferromagnetic Brillouin zone, 
and therefore makes a dominantly negative contribution 
to the Hall conductivity. 
We show in fig. \ref{hallce} an example of temperature dependence 
of the Hall coefficient in electron-doped systems. 
Here $t=0.25$ eV, $t'=-0.45t$, $\mu=-0.146$ eV and $n=1.10$, 
with $\omega_{\rm SF}$ twice as large as that used previously 
for a hole-doped system. 
The Fermi surface is shown in fig. \ref{fs} as the case (c).

\section{Conclusion and Discussions}

It has been pointed out in the literature that 
the remarkable temperature dependence of the Hall coefficient 
in the normal state of the high-$T_c$ cuprates 
is due to 
highly anisotropic scattering of the quasiparticles 
in the presence of strong antiferromagnetic spin fluctuations. 
There is much difference in the damping rate of the quasiparticles 
on the Fermi surface, according to whether a quasiparticle is located 
near the antiferromagnetic Brillouin zone boundary or not. 
However the temperature dependence of the anisotropy 
in the quasiparticle damping is not sufficient to account fully 
for the Hall coefficient. 

In the present paper we have furthermore taken account of 
the vertex correction to the current vertex 
arising from interactions between the quasiparticles. 
We have given a phenomenological expression 
for the dynamical spin susceptibility 
to describe interactions between the quasiparticles. 
In the presence of strong antiferromagnetic spin fluctuations, 
the current vertex is considerably transformed 
from the quasiparticle velocity by the vertex correction, 
and is enhanced at most points on the Fermi surface. 
As a consequence a part of the Fermi surface 
outside of the antiferromagnetic Brillouin zone 
makes negative contribution to the Hall conductivity, 
even if the curvature of the Fermi surface is everywhere hole-like. 
If due regard is taken to the vertex correction to the current vertex, 
the Hall coefficient is much enhanced at low temperatures. 
Temperature dependence of anisotropy in the quasiparticle damping rate 
and of the spin correlation length lead to significant temperature dependence 
of the Hall coefficient. 
In our treatment the Hall conductivity can be negative in electron-doped 
systems, even if the Fermi surface is hole-like. 
A single Fermi surface can make both positive and negative contributions 
to the Hall conductivity, 
and then a sign change of the Hall coefficient 
with a variation of temperature can occur. 
Generally speaking, sign of Hall coefficient cannot be determined 
only by curvature of the Fermi surface 
in systems with highly anisotropic scattering. 

We have adopted a form for the dynamical spin susceptibility which has 
large weight only on the antiferromagnetic spin correlations 
and is not good at accounting for 
the long wavelength component of dynamical spin 
susceptibility\cite{imai,barzykin}. 
Therefore in our approximation scheme for evaluation of the transport current, 
the effect of anisotropic scatterings on the vertex correction 
may be overestimated. 
In actuality the integral equation for the vertex correction no longer 
converges with the antiferromagnetic correlation length $\xi$ 
as large as 5 measured in units of the lattice constant. 
As was discussed below eq. (\ref{vc22}), 
a large $\xi$ restricts regions which make significant contribution 
to the integral equation to those near the ``hot spots'', 
and each ``hot spot'' is coupled to another proportionally to $\xi^2$. 
Then the transport current is very much different 
from the quasiparticle velocity, 
and the integral equation converges slowly.

Our evaluation of the transport current is sensitive 
to shape of the Fermi surface,  
especially to position of the ``hot spots'' as intersection points 
of the Fermi surface with the antiferromagnetic Brillouin zone boundary. 
The ``hot spots'' are strongly coupled with each other, 
and around the ``hot spots'' the Fermi surface is divided into two parts, 
one of which contribute positively and the other of which negatively 
to the Hall conductivity. 
In the present paper we have regarded the Fermi surface for the quasiparticles 
as to be given, and have not selfconsistently determined the Fermi surface 
in the presence of interactions. 
Yanase and Yamada\cite{yanase} and 
the present authors\cite{kontani} 
have pointed out the importance of 
transformation of the Fermi surface due to electron-electron interactions; 
the strong antiferromagnetic spin correlation leads to 
a shape more suitable for nesting and with less curvature. 
From this transformation results a wider range of $T$-linear resistivity. 

In our calculation if we take the ratio $|t'|/t$ larger, 
the ratio of the absolute value of the quasiparticle velocity 
at the ``hot spots'' to that at the ``cold spots'' 
(where $|p_x|=|p_y|$) becomes larger.  
Then effect of the vertex correction gets stronger and the conductivity 
as well as the Hall conductivity becomes enhanced. 
As a result the enhancement of the Hall coefficient is suppressed. 
Therefore, in our treatment, 
less curvature of the Fermi surface leads to    
more enhancement of the Hall coefficient. 
This tendency is likely to be strengthened, if shape of the Fermi surface 
is incorporated to determine characteristics of spin fluctuations. 
The contrary is expected with no regard to the vertex correction 
to the current vertex. 

It has been reported\cite{harris,kubo} 
that, with some generality but being not always the case, 
the Hall coefficient decreases 
as the temperature decreases in a low temperature region. 
We have shown in the previous section that isotropic impurity scattering 
leads to a downturn of the Hall coefficient at low temperatures. 
On the other hand 
Yanase and Yamada\cite{yanase} attributed 
such decrease of the Hall coefficient 
at low temperatures to consequent decrease of the antiferromagnetic 
spin correlation length upon the opening of the pseudogap 
and development of correlations favoring spin singlet states. 
Moreover there exists an intimate 
relationship\cite{ito,takenaka,nakano94,nakano98,mizuhashi,fukuzumi,watanabe}  
between the pseudogap phenomena 
and the transport properties, e.g. the onset temperature of the pseudogap 
$T^*$ is closely related to the beginning 
of the downward deviation of the in-plane resistivity 
from a $T$-linear behavior. 
Therefore the downturn of the Hall coefficient at low temperatures may be 
mainly due to the opening of the pseudogap. 
Also in our theory a decrease of the antiferromagnetic spin correlation length 
leads to a reduction of the Hall coefficient at low temperatures.

In a paper\cite{kontani} in association with the present one, 
an analysis based on 
the fluctuation exchange (FLEX) approximation is given. 
The FLEX is a kind of conserving approximation 
in the sense of Baym and Kadanoff\cite{bickers}. 
In this approximation, characters of spin fluctuations for some 
representative compounds are reproduced. 
With the use of results of the selfconsistent calculations, 
the conductivity and Hall conductivity are calculated in a manner 
which preserves the conservation laws. 
Temperature dependence of the Fermi surface is predicted, and 
this effect 
leads to an temperature dependence of the Hall coefficient 
inconsistent with experiments, 
if the vertex correction to the current vertex is neglected.  
It is shown that the characteristic temperature dependence and 
the enhancement of the Hall coefficient are reproducible only by 
taking account of the vertex correction to the current vertex. 

Lastly we point out that our idea may also account for 
similar behavior in transport phenomena 
of other systems with antiferromagnetic spin fluctuations, 
for example heavy fermion compounds and organic conductors 
such as BEDT-TTF salts. 
To cite a few recent experiments,  
strong temperature dependence of the Hall coefficient 
was reported\cite{nakanishi} 
for a ladder compound Sr$_{14-x}$Ca$_x$Cu$_{24}$O$_{41+\delta}$; 
similarity in transport properties 
with the high-$T_c$ cuprate is discussed 
for a correlated metal V$_{2-y}$O$_3$\cite{rosenbaum}. 

\section*{Acknowledgements}
The authors are grateful to Y. Yanase and Prof. K. Yamada 
for valuable discussions and comments. 


\end{document}